\documentclass[aps,prd,twocolumn,groupedaddress,showpacs,amsmath,amssymb]{revtex4}
\usepackage{graphicx,epsf}
\usepackage[]{latexsym}
\newcommand{\be}{\begin{eqnarray}}
\newcommand{\ee}{\end{eqnarray}}
\newcommand{\bra}[1]{\mbox{$\langle\, #1 \mid$}}

\newcommand{\ket}[1]{\mbox{$\mid #1\,\rangle$}}

\newcommand{\expec}[1]{\mbox{$\langle\, #1\,\rangle$}}

\begin{document}
\title{\bf Short distance signatures in Cosmology:
Why not in Black Holes?}
\author{Roberto Casadio}
\email{casadio@bo.infn.it}
\affiliation{Dipartimento di Fisica, Universit\`a di Bologna, and
I.N.F.N., Sezione di Bologna, via Irnerio~46, 40126 Bologna, Italy}
\author{Laura Mersini}
\email{l.mersini@sns.it}
\affiliation{Scuola Normale Superiore, p.za dei Cavalieri~7,
56100 Pisa, Italy}
%
%
%
\begin{abstract}
Current theoretical investigations seem to indicate the possibility of
observing signatures of short distance physics in the Cosmic Microwave
Background spectrum.
We try to gain a deeper understanding on why all information about this
regime is lost in the case of Black Hole radiation but not necessarily
so in a cosmological setting by using the {\em moving mirror\/} as a
toy model for both backgrounds.
The different responses of the Hawking and Cosmic Microwave Background
spectra to short distance physics are derived in the appropriate limit
when the moving mirror mimics a Black Hole background or an expanding
universe.
The different sensitivities to new physics, displayed by both backgrounds,
are clarified through an averaging prescription that accounts for the
intrinsic uncertainty in their quantum fluctuations.
We then proceed to interpret the physical significance of our findings
for time-dependent backgrounds in the light of nonlocal string theory.
\end{abstract}
\pacs{04.62,+v, 04.70.Dy, 98.80.Cq}
\maketitle
\section{Introduction}
\label{intro}
Although short distance physics belongs in the realm of quantum gravity,
it is possible that some traces may have survived in the low energy
observables.
Many efforts have been devoted to the exploration of this issue
\cite{tp} and the claims fall in two categories:
Hawking radiation \cite{hawking} is {\em robust against\/} nonlinear
modifications of short distance physics \cite{unruh,jacobson,cr}; but
Cosmic Microwave Background (CMB) spectrum may generically be
{\em sensitive\/} to short distance modes \cite{b.greene}
(see also \cite{tp} for a partial list of other relevant References),
except for special classes of modification that ensure adiabatic time
evolution of the modes at late times \cite{mb,branden.jerome}.
\par
These claims should be taken as suggestive rather than conclusive for at
least the two following reasons:
\par\noindent
{\em i\/})
we do not, as yet, have a fundamental theory to describe physics at
energies higher than the Planck mass.
This means that even for specific short distance models introduced in
literature, we do not have a set of equations by which to study these
models.
Einstein equations and in particular the equation of energy conservation
almost always break down in the presence of nonlinear trans-planckian
physics \cite{lm-mb}.
We can hope that quantum field theory and modifications to Einstein
equations which satisfy Bianchi identity may remain a good description,
as approximate tools, while taking the low energy limit;
\par\noindent
{\em ii\/})
the spectrum is very sensitive to the initial conditions.
We do not know what the initial conditions are in Cosmology but they can
contaminate the result of the investigation of the role of nonlinear
physics in the CMB spectrum.
\par
Despite the difficulties of carrying out such an investigation, it is
very exciting to consider the possibility that we may
find low energy signatures from very high energy processes in Cosmology.
If trans-planckian physics is described within the framework of String
Theory (e.g.~see Ref.~\cite{amanda}) with our present state of knowledge,
then its first evidence would be found in cosmological grounds.
\par
Let us take the above claims as true.
In these notes we want to gain a deeper understanding about probes of
Planck scale physics, by posing the following question:
Why do the CMB spectrum and Hawking radiation respond differently to
nonlinear short distance physics?
In order to achieve a comparison between Black Holes and Cosmology,
the setup here is the following: Consider a {\em moving mirror\/} which
follows a certain trajectory along a cartesian direction
(see \cite{birrell} and References therein).
In Section~\ref{mirror} we discuss the response function of a detector
in the space to the right of the moving mirror for two limiting cases:
{\em a\/}) when the mirror's motion ``mimics'' a Black Hole background
(Section~\ref{bhmirror});
{\em b\/}) and when the mirror's motion follows the evolution of the
Hubble horizon (Section~\ref{cosmomirror}).
We then compare the spectra and argue that the difference between Black
Holes and Cosmology with respect to their sensitivity to short distance
physics, arises from the fact that since the {\em stationary\/} background
of Black Holes is characterized by a {\em length scale much larger\/} than
the Planck length, there are lesser degrees of freedom in the first
instance (surface degrees of freedom ``confined'' to the Black Hole
horizon), as compared to the dynamic background of Cosmology which,
due to non-locality, results in volume degrees of freedom.
This leads us on to introducing a spatial averaging prescription to
account for the uncertainty in the origin of Hawking radiation that
indicates a ``wash-out'' of trans-planckian modes, while in Cosmology
there is a natural time averaging to account for the uncertainty in the
initial conditions
which (at best) enhances the role played by short scale physics at later
times.
We conclude in Section~\ref{race} that any traces of Planck scale
physics in the spectra, as measured by their departure from thermality,
may be larger and thus observable in the cosmological case.
We then speculate that our results are related and point very much in
the same direction as recent works on cosmological background
solutions in String Theory (see \cite{eva} and References therein).
An observable effect becomes possible when quantum fluctuations are
comparable to thermal fluctuations.
The competition between quantum and thermal fluctuations is larger for
volume degrees of freedom than for surface degrees of freedom, thus the
possibility that they may be within reach of observation in Cosmology.
\par
Recent work in String Theory \cite{eva} indicates that non-locality may
be required for all non-stationary, cosmologically relevant backgrounds
(i.e.~Lorentzian vacua).
We relate the findings of \cite{eva} to our backgrounds in order to obtain
a physical interpretation for the sensitivity of the CMB spectrum to Planck
scale physics as originating from non-locality of time-dependent solutions
in String Theory.
\par
We use units with $c=\hbar=1$ and denote by $m_p$ the Planck mass and by
$\ell_p=m_p^{-1}$ the Planck length.
Finally, $k_B$ is the Boltzmann constant.
\section{Moving mirrors and the detector}
\setcounter{equation}{0}
\label{mirror}
There are several instances of non-trivial vacua in flat
space-time, such as the time dependent Casimir effect and the
moving mirror, which can be dealt with as time-dependent
boundary conditions for the (Klein-Gordon) wave equation
for a massless scalar field $\phi=\phi(t,x)$:
\begin{subequations}
\be
&&\Box\phi\equiv \ddot\phi-\phi''=0
\label{KG}\\
\nonumber \\
&&\phi(t,z(t))=0
\ ,
\label{left}
\ee
with  $(t,x)$  an (inertial) reference frame~\footnote{We
restrict our considerations to 1+1 Minkowski space-time for the sake
of simplicity.}
and $x=z(t)$ is the trajectory of the moving mirror in such a frame.
We shall assume $z(t<0)=0$, so that the mirror starts moving
at time $t=0$.
For the Casimir effect one also demands
\be
\phi(t,L)=0
\ ,
\label{right}
\ee
\end{subequations}
where $x=L\gg z(t)$ is the location of the second (fixed) wall of
the box~\footnote{With a forethought to modified short distance
physics, one might also trade such sharp boundary conditions for
fuzzier constraints.}.
\par
We note that the case corresponding to the standard moving mirror
\cite{birrell} is then recovered as $L\to\infty$ by assuming that the
space-time extends from $t=-\infty$ (the infinite null past denoted
by $\Im^-$) to $t=+\infty$ (the infinite null future $\Im^+$).
One then chooses appropriate positive frequency modes for the vacuum state
on $\Im^-$ and evolves the system to $\Im^+$, where negative
frequencies and the out-vacuum are well-defined.
This approach is much in the same spirit as the usual in-out
formalism of quantum field theory and
is therefore consistent, provided that the relevant part of the process
occurs in a relatively small portion of the space-time outside
of which the fields propagate freely \cite{fulling}.
The same approach is taken for a Black Hole background and in fact,
one finds out that for a particular choice of the mirror's trajectory,
the moving mirror case reproduces Hawking radiation, {\em when
backreaction is ignored\/}~\footnote{For an attempt to extend the
analogy when the backreaction is included see Ref.~\cite{cv2}.}.
Both processes can be viewed as a projection from the surface
$\Im^-$ to the mirror (horizon) and then back to the surface
$\Im^+$, which points to the holographic nature of the effect.
\par
A more complex approach which fully incorporates the backreaction
of particles on the background geometry \cite{blhu} appears more
appropriate for describing cosmological particle creation.
In this case one can assume the space-time begins at $t=0$
(corresponding, e.g.~to the onset of inflation).
Positive frequency modes can correspondingly be defined only with
respect to the initial time surface at $t=0$ and the initial vacuum
evolved according to the in-in, or closed time path formalism of
quantum field theory \cite{ctp}.
The notion of particles is not unique in this fully time-dependent
context \cite{fulling}, although one may be able to define it if a
``preferred'' frame exists.
However, no projection of the sort described in the previous paragraph
is naturally conceivable because of the arbitrary time dependence
of the mirror (cosmological scale factor).
\subsection{Casimir effect and Response function for the
``Black Hole'' mirror}
\label{bhmirror}
The analogy between Hawking radiation from a Black Hole and a moving
mirror is well known.
Here we shall simply recall the main points reported
in~Ref.~\cite{birrell}, chap.~4.4.
Let us introduce the usual advanced and retarded coordinates,
$u=t-x$ and $v=t+x$, which cover all of Minkowski space-time to the
right of the mirror.
The mirror trajectory is described by $x=z(t)$.
A convenient choice of modes solving Eqs.~(\ref{KG}) and (\ref{left})
is then given by
\be
\phi_\omega^{\rm in}(u,v)={i\over\sqrt{4\,\pi\,\omega}}\,
\left[e^{-i\,\omega\,v}-e^{-i\,\omega\,(2\,t_u-u)}
\right]
\ ,
\label{bhm}
\ee
where $t_u=u+z(t_u)$.
Such modes are positive frequency on the infinite null past ($\Im^-$
or $u=-\infty$) and define the in-vacuum $\ket{0_{\rm in}}$ which,
for $t<0$, remains devoid of particles (since $t_u=u$ for $u<0$).
Modes of negative frequency can analogously be introduced on $\Im^+$
(i.e.~$v=+\infty$).
\par
We are interested in finding the squeezing of the final state at $\Im^+$
or $u=+\infty$ as compared to the initial vacua, which results from
the mirror's motion.
When the mirror starts moving at $t=0$, the in-vacuum is modified by
the Doppler factor appearing in the right-moving part of the modes
(\ref{bhm}).
The case which mimics Hawking radiation is given by a mirror's trajectory
that approaches the speed of light,
\be
z(t)=-t+v_0\,\left(1-e^{-2\,\kappa\,t}\right)
\ .
\ee
The last term above represents a transient between the steady state
and the asymptotic null curve $v=v_0$.
Left-moving parts of modes (\ref{bhm}) get reflected off the mirror
for $v\le v_0$, while they pass unaffected for $v>v_0$.
The piling up of (equispaced) lines of constant $u$ along the
``causal horizon'' $v=v_0$ occurs also for a collapsing
body and is at the origin of the effect (see Refs.~\cite{visser} for
a neat exposition).
\par
A detector can now be introduced as a quantum system with (discrete)
internal energy spectrum $\ket{E_n}$ and (sharply) localized on the
trajectory $x=y(t)$.
It interacts with the scalar field via a dipole term,
\be
V_{\rm int}=q\,\delta\left(x-y(t)\right)\,\hat Q\,\hat\phi(t,x)
\ ,
\label{Vint}
\ee
where $q$ is a (small) coupling constant and $\hat Q$ the operator
which causes transitions between inner energy levels.
The response function of such a detector is just sensitive to positive
frequencies of the scalar field (to lowest order in $q$), and its
probability of excitation from the ground state $\ket{E_0}$ to the
state with energy $E_n$ is given by \cite{birrell}
\begin{widetext}
\be
P_{0\to n}\!\!\!&=&\!\!
q^2\left|\bra{E_n}\,\hat Q\,\ket{E_0}\right|^2
\int dt\,dt'\,e^{-i\,(E_n-E_0)\,(t-t')}
\int dx\,dx'\,\delta(x-y(t))\,\delta(x'-y(t'))\,
\bra{0_{\rm in}}\,\hat\phi(t,x))\,\hat\phi(t',x'))\,
\ket{0_{\rm in}}
\nonumber \\
\!\!&=&\!\!
q^2\,\left|\bra{E_n}\,\hat Q\,\ket{E_0}\right|^2\,
\int dt\,dt'\,e^{-i\,(E_n-E_0)\,(t-t')}\,
\bra{0_{\rm in}}\,\hat\phi(t,y(t))\,\hat\phi(t',y(t'))\,
\ket{0_{\rm in}}
\ .
\label{inte}
\ee
\end{widetext}
The parameter $\kappa$ entering in the expression of
the Wightman function for the in-vacuum appears in the above
integral in such a way that the detector responds as if it is
immersed in a thermal bath,
\be
P_{0\to n}=
q^2\,{\left|\bra{E_n}\,\hat Q\,\ket{E_0}\right|^2\over E_n-E_0}\,
{1\over e^{(E_n-E_0)/k_B\,T}-1}
\ ,
\label{thermo}
\ee
with Doppler shifted temperature
\be
k_B\,T={\kappa\over 2\,\pi}\,\sqrt{{1-w\over 1+w}}
\equiv{\kappa\over 2\,\pi}\,\Gamma(w)
\ ,
\label{doppler}
\ee
where $w$ is the (instantaneous) velocity of the detector in the
$(t,x)$ frame, $y(t)\simeq y(0)+w\,t$.
\par
The above simple result follows from the spatial $\delta$-function
in Eq.~(\ref{Vint}) forcing the Wightman function to be strictly
evaluated on the trajectory's points $y(t)$ and $y(t')$
{\em without\/} smearing~\footnote{Let us note that such a
$\delta$-function implies a sum over all normal modes of the
detector's position wave-function which would also be affected
by modifications in the trans-planckian regime.
We gloss over this fact here precisely for the reason presently
explained.
\label{delta}}.
A realistic detector would however be better described in terms
of a wave-packet $\Psi(t,x)$ peaked along the trajectory $x=y(t)$
with a ``width'' $\Delta$.
Hence, the $\delta$-function in Eqs.~(\ref{Vint}) and (\ref{inte})
must be replaced by $|\Psi(t,x)|^2$ and one can show that scalar
field modes with short wavelengths $\lambda\ll\Delta$ are averaged over
and do not contribute appreciably to the transition probability
(\ref{inte}) (for the details see, e.g.~\cite{cv} and
References therein).
One can understand the origin of this suppression by considering a
region of size $\Delta\sim N\,\lambda$ (with $N\gg 1$) and estimating
the order of magnitude in fluctuations therein by using the uncertainty
principle for the location of the $\Delta$ region.
Assuming a Poisson distribution, one finds
\be
{\delta \Delta\over\Delta}
\sim {\delta N\over N}
\sim {\sqrt{N}\over N}
\sim \sqrt{\lambda\over\Delta}
\ll 1
\ .
\label{poisson}
\ee
For a realistic detector $\Delta\gg\lambda=\ell_p$, thus it would
never respond to trans-planckian modes and the result (\ref{thermo})
holds with very good approximation for sufficiently low (detected)
temperature $k_B\,T\lesssim\Delta^{-1}$.
\par
It is futile to ask where those thermal particles
come from since the very notion of a particle is likely lost near
the source region \cite{fulling}.
One could say that particle production roughly occurs in the strip
$0<v<v_0$ where the mirror changes its motion between the two
asymptotic regimes.
However, detection is performed well outside the strip $0<v<v_0$,
and most commonly at very late times (near $\Im^+$), where a
particle interpretation is recovered.
In any case, since the typical wavelengths of the produced quanta are
of order $\kappa^{-1}$, a stationary detector (with $w=0$)
cannot resolve the position of the source with accuracy better then
$\kappa ^{-1}$.
There is thus one fundamental and intrinsic (i.e.~detector independent)
length scale in the effect, namely $\ell\sim\kappa^{-1}$, which
determines the statistical mechanics of the emitted flux as well as
the source fuzziness.
The other length $v_0$ reflects an irrelevant detail of the mirror
motion, while $w$ accounts for the relative state of the detector with
respect to the mirror.
\par
Let us now see how this picture translates into Black Hole language.
We recall that for a Schwarzschild Black Hole of Arnowitt Deser Misner
(ADM) mass parameter $M$ the Hawking temperature is
\be
T_H={1\over 8\,\pi\,M\,k_B}
\ ,
\ee
and the Tolman factor for a stationary detector placed at $r\gg 2\,M$
which receives signals from a source located at radial position $r=r_s$
is
\be
\Gamma_T=\sqrt{1-{2\,M\over r_s}}
\ .
\label{tfac}
\ee
Hence, if the particles came precisely from $r_s=2\,M$, they had to be
produced with a local energy $k_B\,T\to\infty$ [achieved by taking
$w\to-1$ in Eq.~(\ref{doppler})].
One can then say that $k_B\,T$ in Eq.~(\ref{doppler}) is to be viewed
as the energy of Hawking quanta at the point of emission with
$\kappa=2\,\pi\,k_B\,T_H=1/4\,M$ the surface gravity of the Black Hole,
and that the Doppler factor $\Gamma(w)$ plays the role of the Tolman
factor, with $w$ a function of the radius $r_s$, the origin of the
radiation.
Since $\kappa\ll m_p$ (as $M\gg \ell_p$ for a Black Hole), the Hawking
temperature $T_H$ is far less than the Planck mass and trans-planckian
energies will not reach a stationary detector (which could anyways
not detect them according to the uncertainty principle argument above
since its width $\Delta\gg \ell_p$).
\par
The problem remains whether modes of trans-planckian energies are
excited at some small radius $r_s\sim 2\,M$ for which $k_B\,T\gtrsim m_p$
\cite{amanda}.
Since Hawking quanta have typical wavelengths of order $M$,
a detector cannot resolve their origin with an accuracy better than
$M$ \cite{scardigli} (modulo the Tolman factor) as in the mirror
case.
Further, numerical studies of wave propagation around ``sonic'' Black
Holes show that outgoing modes are excited as a non-adiabatic effect
when their wavelengths become comparable to the typical length scale
$M$ of the change of the background curvature (gravitational potential),
thus this occurs sufficiently far away from the horizon \cite{unruh}.
One can therefore roughly consider the Black Hole as an
emitter~\footnote{Also notice that due to the spherical symmetry and
the stationary Black Hole background this case becomes one dimensional
in terms of fluctuations.}
of size $\Delta_{BH}\sim M\sim N\,\ell_p$ and repeat the Poisson
counting which led to Eq.~(\ref{poisson}).
One obtains an order of magnitude estimate for fluctuations inside
the region of size $\Delta_{BH}$,
\be
{\delta \Delta_{BH}\over\Delta_{BH}}
\sim \sqrt{\ell_p\over M}
\ll 1
\ ,
\ee
This order of magnitude estimate supports the conclusion that
trans-planckian fluctuations are suppressed from the onset.
\par
The above argument can be refined by introducing an effective Tolman
factor (for the detector far away from the Black Hole) which takes
into account the uncertainty in the radius of the source $r_s$ and
consequently of the temperature $T$ of the emitted Hawking radiation.
This is obtained by averaging the Doppler factor $\Gamma$ in
Eq.~(\ref{doppler}) over a corresponding range of equivalent source
positions to yield the average energy at the point of emission
corresponding to a given detected $T_H$,
\be
\rho\equiv
{\expec{k_B\,T}\over k_B\,T_H}
={1\over \ell}\,\int_0^\ell dx\,\sqrt{{1-w(x)\over 1+w(x)}}
\ .
\label{tfact}
\ee
Let us then replace the Doppler factor with the Tolman factor
(\ref{tfac}) and assume $\ell=\alpha/\expec{k_B\,T}$,
with $\alpha\sim 1$.
Upon substituting in Eq.~(\ref{tfact}), one obtains
\be
\rho
&=&
{\expec{k_B\,T}\over\alpha}\,\int_{2M}^{2M+\alpha/\expec{k_B\,T}}
{dr_s\over\sqrt{1-{2\,M/r_s}}}
\nonumber \\
&=&
{\rho\over 4\,\pi\,\alpha}\,
\int_0^{\alpha\over 4\,\pi\,\rho}dx\,\sqrt{1+x^{-1}}
\ .
\ee
The above expression requires the self-consistency condition
for $\rho$, {\em independent\/} of $M$,
\be
{1\over 4\,\pi\,\alpha}\,
\int_0^{\alpha\over 4\,\pi\,\rho}dx\,\sqrt{1+x^{-1}}
=1
\ .
\label{self}
\ee
The (numerical) solution for $\alpha=1$ is
\be
\expec{k_B\,T}\simeq 1.23\,k_B\,T_H
\ ,
\label{y}
\ee
which is much smaller than $m_p$ for $M\gg\ell_p$.
This result clarifies the fact that high energy modes
``wash-out due to averaging''.
\par
Let us note that, since the Tolman factor $\Gamma_T$ does
not depend on the energy $k_B\,T_H$, neither does the ratio $\rho$,
and the result (\ref{y}) holds for each frequency of the
Hawking spectrum separately.
Finally, for $\ell\ll M$ ($\alpha\ll 1$) one would recover the
usual result $k_B\,T\to \infty$.
\subsection{Casimir effect and Response function for the ``cosmological''
mirror.}
\label{cosmomirror}
The CMB spectrum in Cosmology is formally calculated by the same methods
as particle creation in curved space \cite{branden.jerome}, since the
nonlinear time-dependent frequency of short distance physics can be
attributed to a curved space time geometry \cite{mb}.
In our case this role is played by the
``cosmological mirror's trajectory''.
In this part we will follow a different approach from the conventional
one in calculating the Bogolubov coefficients.
Instead of finding the S-matrix element for the asymptotic vacua at
past and future null infinity, the time-dependent Bogolubov coefficients
are calculated iteratively at each time slice that is taken consecutively
after small time intervals.
The purpose for following this procedure is in order to enable us
to trace step by step the stage at which the effects of trans-planckian
signatures may arise.
Recall that unlike the ADM mass parameter $M$ in Black Holes there is
{\em no characteristic time scale in Cosmology} besides the gravity scale
(the Planck length $\ell_p$), thus the time step can be taken as small
as $\ell_p$.
\par
Let us define the initial Fock space from positive frequency modes at
$t=0$ and then evolve to larger values of $t$.
In doing so, we are assuming that the initial conditions are known and
well defined.
In particular, we note that solutions to Eq.~(\ref{KG}) are given by
superposition of plane waves in the coordinates $x$ and $t$.
Then imposing Eqs.~(\ref{left}) and (\ref{right}) yields the
(discrete) spectrum
\be
\phi_n(t,x)={\mathcal N}\,e^{-i\,\omega_n\,t}\,
\sin\left(n\,\pi\,{x-z\over L-z}\right)
\ ,
\ee
where $n$ is a positive integer and ${\mathcal N}$ is a normalization
factor.
The frequency $\omega$ equals the spatial momentum $k$ and is given
by
\be
\omega_n=k_n={n\,\pi\over L-z}
\ .
\label{modes}
\ee
The modes (\ref{modes}) indeed solve Eq.~(\ref{KG})
only if one can neglect the time dependence of $z$ (and, consequently,
in $\omega$).
This is basically the lowest order adiabatic approximation, which helps in
defining the concept of particles \cite{fulling}.
We shall provide later a way of measuring its validity.
\par
In order to build up an Hilbert space one introduces the usual (fixed
time) scalar product
\be
(\phi,\psi)\equiv-i\,\int_z^L dx\,
\left(\phi\,\partial_t\psi^*-\psi^*\,\partial_t\phi\right)
\ ,
\label{scal}
\ee
which yields the (time independent) result
$(\phi_n,\phi_m)=\delta_{n,m}$,
provided the normalization is ${\mathcal N}=1/\sqrt{n\,\pi}$
and the time dependence of $z$ is neglected.
A general instantaneous solution can therefore be written as
\be
\phi(t,x)=\sum_{n=1}^\infty\,
\left[a_n\,\phi_n(t,x)+a_n^\dagger\,\phi_n^*(t,x)\right]
\ ,
\label{exp}
\ee
where $a_n$ and $a_n^\dagger$ become the annihilation and creation
operators of the quantized theory \cite{birrell}.
\par
To take fully into account the time dependence of $z$ one should
consider time-dependent coefficients $a_n=a_n(t)$ and substitute
the expression (\ref{exp}) into Eq.~(\ref{KG}).
One would then find that $a\sim \delta t/L\equiv\epsilon$, where
$\delta t$ ($\ll L$) is the typical time step
considered (see below).
Therefore $\epsilon$ is the parameter which measures the departure
from adiabaticity and the time dependence of $a$ yields second order
effects in $\epsilon$.
A systematic treatment consisting in a perturbative expansion in
$\epsilon$ can be carried out \cite{lewis}, but does not seem
necessary for the treatment of the conceptual issue investigated
here.
\par
The instantaneous vacuum at $t=t_1$ is defined as usual by
$\hat a_n\,\ket{0(t_1),t_1}=0$.
At a different time $t=t_2>t_1$ one has a different set of modes and
consequently a different expansion of the form (\ref{exp}).
This also means a different vacuum $\ket{0(t_2),t_2}$ or, equivalently,
that the state $\ket{0(t_1),t_2}$ will generically contain
``particles''.
Due to the time dependence of the background, the calculation of the
spectrum in this scenario reduces to the familiar problem of particle
creation in curved space \cite{parker,hawking,birrell,fulling}.
\par
\begin{figure}
\raisebox{3cm}{$\delta N_n$}
\epsfxsize=2.9in
\epsfbox{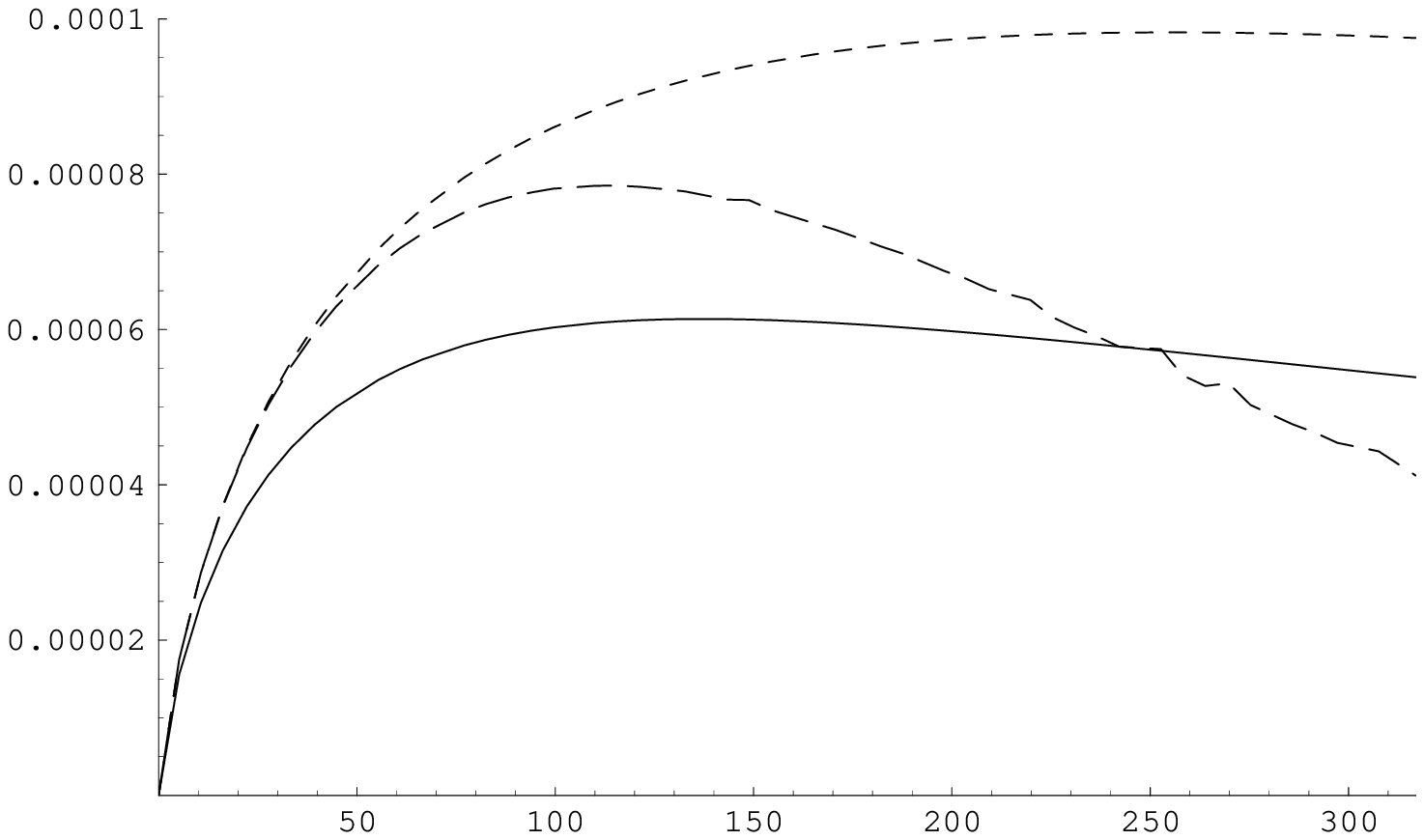}\\
\hspace{-0.2in}
\raisebox{0.5cm}
{\hspace{6.5cm}$n$}
\caption{Increase in the number of particles $\delta N_n$
[as in Eq.~(\ref{dN})] for
$\gamma=10^{-3}$ and linear dispersion relation with (continuous line)
and without (dashed line) a sharp cut-off at $\omega_n=m_p$ ($n\sim 300$).
The longer dashed line is obtained from the Epstein dispersion
relation (\ref{ep}).}
\label{fig1}
\end{figure}

Below we denote by unbarred symbols quantities evaluated
at $t=t_1$ and with a bar the same quantities at the slightly later
time $t_2=t_1+\delta t$ (with $\delta t\ll t$).
Further, the ``cosmological mirror position'' at $t=t_2$ is given
by $\bar z=z+\delta z$
($|\delta z|\simeq|\dot z|\,\delta t\le\delta t\ll L$).
The increase $\delta N_n$ in the ``number of particles'' in the mode $n$
stored in the state $\ket{0(t_1),t_1+\delta t}$ is obtained from the
Bogolubov coefficients
\begin{widetext}
\begin{subequations}
\be
\beta_{nm}\equiv (\bar\phi_n,\phi_m^*)
&=&e^{-i\,(\omega_m+\bar\omega_n)\,\delta t}\,
{\omega_m-\bar\omega_n\over\pi\,\sqrt{n\,m}}\,
\int_{z+\delta z}^L dx\,
\sin\left[\bar k_n\,(x-z-\delta z)\right]\,\sin\left[k_m\,(x-z)\right]
\label{betaa}
\\
&=&
{e^{-i\,(\omega_m+\bar\omega_n)\,\delta t}\over\pi\,\sqrt{n\,m}}\,
{(L-z)\,n\over (L-z)\,n+(L-z-\delta z)\,m}\,
\sin\left({m\,\pi\,\delta z\over L-z}\right)
\ ,
\label{beta}
\ee
\end{subequations}
\end{widetext}
and is given by
\be
\delta N_n=\sum_{m=1}^\infty\,\left|\beta_{nm}\right|^2
=\sum_m\,{n\over\pi^2\,m}\,
{\sin^2\left(m\,\pi\,\gamma\right)\over
\left[n+(1-\gamma)\,m\right]^2}
\ ,
\label{dN}
\ee
with $\gamma\equiv\delta z/(L-z)\sim \epsilon\ll 1$.
An example of $\delta N_n$ for $\gamma\sim 10^{-3}$ is plotted in
Fig.~\ref{fig1}.
It is interesting to note that in the adiabatic approximation of a slowly
varying $z$, the shape of the curve resembles a planckian
distribution.
Decreasing $\gamma$ lowers the peak and shifts it to larger values
of $n$.
\par
For $\delta z\to 0$ (stationary mirror) and/or
$L-z\to\infty$, $\gamma\to 0$, each term in the sum vanishes,
hence $\delta N_n\to 0$.
However, for $L$ finite, the total number of particles $N_n$ in
the mode $n$ after a (large) lapse of time $T$ is obtained from
the expression in Eq.~(\ref{dN}) summing over a large number
($T/\delta t$) of small steps $\delta t$.
This would yield $N_n\sim (T/\delta t)\,\delta N_n$.
Eventually $T\to \infty$ and it is therefore clear that
$0=\lim_{T\to\infty}\left(\lim_{L\to\infty}\,N_n\right)\not=
\lim_{L\to\infty}\left(\lim_{T\to\infty}\,N_n\right)=\infty$.
\par
\begin{figure}
\raisebox{3cm}{$\delta N_k$}
\epsfxsize=2.8in
\epsfbox{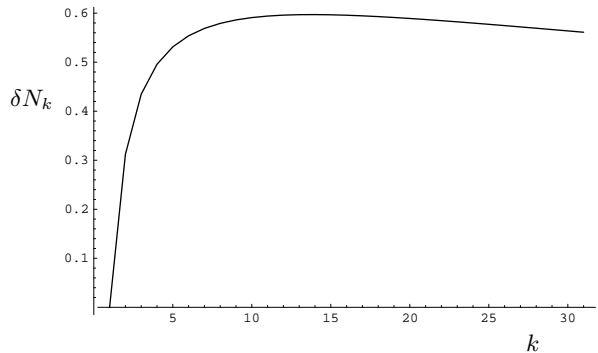}\\
\hspace{-0.2in}
\raisebox{0.5cm}
{\hspace{6.5cm}$k$}
\caption{Increase in the particle density $\delta N_k$
[as in Eq.~(\ref{dNk})] for $\delta z=1$.}
\label{fig2}
\end{figure}

The wall at $x=L$ can also be removed by relaxing the condition
(\ref{right}) from the very beginning which then results in a
continuous spectrum of states (as distributions)
\be
\phi_k(t,x)={{\mathcal N}\over\sqrt{\omega}}\,e^{-i\,\omega\,t}\,
\sin\left[k\,\left(x-z\right)\right]
\ ,
\ee
where $\omega$ is a positive real and ${\mathcal N}$ the
normalization factor in the scalar product (\ref{scal}).
It is now possible to repeat the same steps as before and,
on observing that
\be
&&\int_{z+\delta z}^\infty dx\,
\sin\left[k\,(x-z-\delta z)\right]\,\sin\left[q\,(x-z)\right]
\nonumber \\
&&={k\over k^2-q^2}\,\sin\left(q\,\delta z\right)
\ ,
\ee
one obtains (modulo numerical normalization factors) a spatial
density of produced particles
\be
\delta N_k\simeq\int_0^\infty {dq\over \sqrt{q}}\,
{k^2\over\left(k+q\right)^2}\,\sin^2\left(q\,\delta z\right)
\ .
\label{dNk}
\ee
The above expression has the same features as that in Eq.~(\ref{dN})
for the discrete case:
the peak in the distribution lowers and shifts to larger values
of $k$ for decreasing $\delta z$
(see Fig.~\ref{fig2} for an example).
The total density of particles produced in the mode $k$ would
then be given by
\be
N_k=\sum_{\delta z} \delta N_k
\ ,
\ee
where the sum is over the (discretized) displacements $\delta z(t)$
of the moving mirror.
For a mirror which approaches the speed of light $\delta z\sim 1$
and one finds
\be
N_k(T)\sim T\,\delta N_k
\ ,
\ee
assuming the motion has started at $t=0$ and ended at $t=T$, and
$\delta N_k$ is that plotted in Fig.~\ref{fig2}.
\par
It is now straightforward to study the consequences of non-linear
dispersion relations
\be
\omega_n=\omega(k_n)\not=k_n
\ .
\ee
In fact, again neglecting the time dependence of $z$, the only
modification that ``registers'' parameters of short distance
non-linearity occurs in the factor in front of the integral in
Eq.~(\ref{betaa}).
For example, one can simply place a sharp cut-off at $\omega=m_p$
\cite {kempfniemeyer} or consider the smooth Epstein dispersion
relations of Ref.~\cite{epstein},
\be
\omega_n\sim k_n\,{\rm sech}\left({k_n\over\mu}\right)
\ ,
\label{ep}
\ee
(where $\mu\sim 1$ determines the location of the peak in the spectrum),
to obtain the (generally large) damping of particle production displayed
in Fig.~\ref{fig1}.
Although the high wave-number effect can be directly appreciated at the
time of production only for modes with wavelengths sufficiently large
(i.e.~$n$ sufficiently small),
the cosmological red-shift will make the entire region in Figs.~\ref{fig1}
and \ref{fig2} (and possibly a larger one) visible to present detection,
as we now proceed to discuss.
Therefore this effect may become more important at present than during
inflation.
\par
In order to translate the Casimir results into Cosmology
we first need to put down the dictionary.
One can consider the size of the box $L-z$ as the scale factor
of the Universe $A(t)$ whose evolution is a priori not known.
Having this in mind the mirror's trajectory $z(t)$ was kept
arbitrary, the evolution of the system followed continuously
(and not just looking at the asymptotic behavior of the mirror)
and particle production was seen to occur homogeneously.
For sufficiently smooth trajectories (i.e.~cosmological evolution),
a stationary ($w=0$) detector would measure a portion
$0<k\lesssim\Delta^{-1}$ of the instantaneous spectrum shown in
Fig.~\ref{fig1} and \ref{fig2} at the time $t_s$ at which particle
production occurs.
Energies are successively red-shifted by the expansion of the
Universe and the same detector probes the portion
$0<k\lesssim\Gamma(w)\,\Delta^{-1}$ at the time $t\gg t_s$,
where $\Gamma$ is the Doppler factor of Eq.~(\ref{doppler}) with $w$
a function of the time of creation $t_s$.
If the cosmological evolution is such that $w\to -1$ for $t_s\to 0$
(considered, e.g.~as the onset of inflation), trans-planckian modes
certainly appear in the CMB spectrum at the late time $t$.
\par
For the Black Hole case the uncertainty in the location of the
origin of Hawking quanta led us to introduce a spatial average
over the region of size $\ell\sim \kappa^{-1}\gg\ell_p$ outside the
horizon [Eq.~(\ref{tfact})] by which we have shown that
short distance physics does not in any way affect the statistical
mechanics of the detected Hawking spectrum (see also
Refs.~\cite{unruh,jacobson,cr}).
In the cosmological case, the significant uncertainty is in the
creation time $t_s$ of cosmological fluctuations for a given
detected energy at time $t$.
One could thus envisage a temporal averaging over such a time
uncertainty.
However, no such procedure could prevent the high frequency branch
of the spectrum to become visible in an inflationary scenario
\cite{branden.jerome}.
In fact, suppose the duration of the de~Sitter phase, with
scale factor
\be
A\sim e^{H\,t}
\ ,
\ee
is $T\pm\Delta T$ with uncertainty $\Delta T>0$,
then an effective cosmological blue-shift would be given by
\be
\expec{\Gamma}&\sim&
{1\over2\,\Delta T}\,\int_{+\Delta T}^{-\Delta T}
dt_s\,\left({e^{H\,T}\over e^{H\,t_s}}-1\right)
\nonumber \\
&=&{e^{H\,T}\over H\,\Delta T}\,\sinh(H\,\Delta T)-1
>e^{H\,T}
\ ,
\ee
which, for all uncertainties $\Delta T>0$, results in a larger
blue-shift than one would have for $\Delta T=0$.
\par
We can conclude that the most important difference between the Hawking
effect and the CMB spectrum in Cosmology is that, due to their
stationarity, Black Holes have a fixed characteristic length scale
$\ell\sim M\gg \ell_p$ which sets the size of relevant fluctuations
well above the Planck scale.
In Cosmology the analog scale is the inverse of the {\em time-dependent\/}
Hubble constant $H=\dot A/A$ whose value varies in time and, at best,
is just bound from below by the natural quantum gravity scale $\ell_p$.
Quantum fluctuations in a cosmological setting would therefore
be expected to correspond to Planck scales.
This is to say that according to the uncertainty principle, the
Hubble radius $H^{-1}$ is uncertain within (roughly) one Planck
length.
As discussed in the next Section, the interpretation of this statement
in terms of a realistic cosmological scenario of Lorentzian vacua given
by String Theory in time-dependent backgrounds, is that non-locality
seems to be required for at least Planck length scales
\cite{eva}~\footnote{However, the validity of these statements and the
uncertainty principle near the Planck scale should be treated with caution
since we have no knowledge of the physics at these energies.
Besides space-time non-locality raises important doubts for the
unitarity of the theory at these scales,
perhaps stochastic processes might best describe physics in this regime
and they may smear out any relevant information of new physics
\cite{blhu}.}.
\par
The authors in Refs.~\cite{kempfniemeyer,padmanabhan} used similar quantum
statistical mechanics arguments to get an order of magnitude estimate
of the contributions of Planck scale fluctuations in Cosmology and
found it to be of the order $(H\,\ell_p)^2$.
Other works estimate the imprints of Planck scale physics to be of
the order $(H\,\ell_p)$ \cite{b.greene}.
This issue is still under
debate due to the ambiguity in the definition of initial conditions.
However the important point for our investigation is that if, besides
the quantum gravity scale $\ell_p$, the only other characteristic scale
of the cosmological system is the inverse Hubble constant $H^{-1}$
which, due to its time dependence, has a lower bound of order $\ell_p$,
then observational signatures of new physics at Planck scale are not
far out of reach.
\par
Let us illustrate with a concrete example, by considering the response
function of a detector for a mirror's trajectory $z(t)$, whose motion
in the intermediate regime resembles the nonlinear short distance
physics of a Bianchi type I Cosmology \cite{birrell} (with the
anisotropic parameters $\delta_i=\beta=0$).
Below we have to take into account that the time dependence in $z(t)$
given by the parameter $\alpha_k$ and Eq.~(\ref{inte}), when combined
with the Bogolubov coefficients (\ref{betaa}), becomes a non-vanishing
complicated function of the initial vacuum $\ket{0(t_s),t}$ that
registers parameters of short distance physics contained in $z(t)$
and does not reduce to the thermal distribution of Eq.~(\ref{thermo}).
In this example the generalized frequency term that enters Eq.~(\ref{KG}),
for modes with wavenumber $k$, is given by
\be
\omega^2=k^2\,\left(1-e^{-\alpha_k\,\eta^2}\right)
\ ,
\ee
with $\eta$ the conformal time.
The (diagonal) Bogolubov coefficients are \cite{birrell}
\be
\beta_{kk}=
-\frac{i\,\pi^{1/2}}{2\,\omega\,\sqrt{\alpha_k}}\,
\exp\left({2\,\omega\over\alpha_k}\right)
\ .
\ee
The expression for $\beta_{kk}$, which enters in the response function
and registers the short distance parameter $\alpha_k$, illustrates that
unlike the Black Hole case where degrees of freedom are ``projected''
into its surface and thus any traces of Planck scale physics are lost
due to this confinement to the horizon, in Cosmology no such confinement
occurs.
The detector registers details of the mirror's motion ($\alpha_k$)
and the response function is generally time dependent.
In this case the detector is registering volume degrees of freedom due
to non-locality of the time-dependent space-time.
\section{The real race: Black holes versus Cosmology}
\label{race}
Let us try below to interpret our findings of Section~\ref{mirror}
and search for a better conceptual understanding for the display of
different sensitivities of Black Holes and Cosmology to Planck scale
physics.
In doing so, we rely on the assumption that String Theory \cite{gsw}
is the viable candidate for describing very high energy physics for
both cases.
We also speculate on the relation of the sensitivities of these
backgrounds to a Holographic interpretation \cite{holo} where
possible.
\par
From a string theoretical point of view, non-locality seems to be required
for time-dependent backgrounds (see \cite{eva} and References therein),
while locality is sufficient for describing stationary solutions.
Despite much of the current efforts in literature, the Holographic
Principle is not yet well understood or known for most of the realistic
cases of time-dependent boundaries.
It is possible that the difficulty of a holographic interpretation for
a realistic cosmological scenario may be originating from the requirement
of non-locality of String Theory actions, arising in time-dependent
boundaries.
For this reason, the projection of the degrees of freedom on the
surface might not be possible due to long-range correlations
produced by non-local terms on the world-sheet action.
This might be why in Section~\ref{mirror} the sensitivity of
the radiation spectrum to short scale physics, for a {\em stationary\/}
versus a {\em dynamic\/} scenario turned out equivalent to having surface
projection versus volume degrees of freedom in the latter.
Non-locality introduces long-range correlations and therefore it does
not allow a projection of the degrees of freedom to the boundary.
\par
Time-dependent string backgrounds solutions, relevant for Cosmology,
were studied in \cite{eva} for a special class of geometries
of particular interest because, although not an exact CFT, they are
still solutions to Einstein equations.
The results of Ref.~\cite{eva} seem to indicate that within the framework
of perturbative String Theory, a squeezed state on the boundary
(equivalent to having particle creation there) is generically derived
from nonlocal actions on the boundary which are obtained from vertex
interactions that deform the string world-sheet.
However, conformal invariance on the world-sheet is preserved when
string perturbation theory is valid but, {\em Nonlocal String Theory
seems to be required to describe Lorentzian vacua\/} since in dynamic
backgrounds the Bogolubov transformation is nontrivial.
Here, our understanding of the discrepancy between the Planck physics
sensitivity of the Hawking radiation and CMB spectrum due to the volume
degrees of freedom for non-stationary backgrounds, is very much in
accord with the findings of Ref.~\cite{eva} for the necessity of
non-locality for time-dependent backgrounds.
\par
In Section~\ref{mirror} we showed through the moving mirror that
for a ``constant'' Black Hole mass we have a projection and confinement
of the relevant information from the surface $\Im^-$ to the (smeared)
horizon of the Black Hole.
Although the results for the moving mirror that reproduces Hawking
radiation are well known, our interpretation of the results by comparison
to Cosmology is new.
In order to inquire about the possible departure of the spectrum from
thermality due to the new physics at short distances, we needed to
compare quantum fluctuations to thermal fluctuations.
Assuming that the uncertainty principle is valid at least up to Planck
scale, we used it as a tool in order to show that since the
Hawking radiation from a Black Hole (whose horizon radius $M\gg \ell_p$)
has typical wavelengths many orders of magnitude larger than the Planck
length, then Planck scale physics becomes irrelevant while averaging over
such a ``large region''.
Much in the same spirit the Holographic Principle provides an explanation
for the Black Hole's degrees of freedom being projected to its fixed
boundary, the horizon \cite{holo}.
Mathematically this projection of the degrees of freedom to the surface
becomes clear from the $\delta$-function that appears in the response
function of the detector (see Section~\ref{bhmirror}).
It therefore seems that the memory loss that the spectrum suffers in terms
of the Planck scale physics comes from the reduction of the degrees of
freedom to the surface of the Black Hole, which results from
stationary backgrounds with a fixed characteristic scale $M\gg\ell_p$.
In this case, one can understand the reduction of the Black Hole's degrees
of freedom as arising from two principles:
holography and uncertainty, which are both valid for
{\em stationary boundaries\/} of  size larger than the Planck length.
\par
We would naively expect a similar situation to occur in Cosmology.
In Cosmology the number of e-foldings, $N=\ln(A)$, is given by the Hubble
constant $H$ (where $H=\dot N$ and $A$ is the scale factor).
Thus $H$ would be the parameter equivalent to the ADM mass of the Black
Hole since they both determine the red-shift (of radiation
respectively propagating forward in time or approaching the Black Hole
horizon).
From Section~\ref{cosmomirror} we see that the detector will register,
besides its non-inertial motion with respect to the comoving frame with
velocity $w=w(H)$, also the nonlinear deviations of short scale physics
(e.g.~Bianchi type~I in \cite{birrell} can be our imperfect fluid) and the
nonlinear deviations from the Hubble linear law.
As we know, in our universe $H$ has changed tremendously with time,
say between $z=10^{60}$ inflation era, to $z=10^3$ the red-shift of the
last scattering surface.
The only instance when we can obtain an analogous situation with the
Black Hole case for the detector, is if we have a perfectly linear Hubble
law $z=H\,s$, with $H$ constant, $s$ the proper distance and the
temperature is rescaled as $T=A\,T$ in the comoving frame.
Then a similar situation to the Black Hole should arise, provided that
our detector is moving non-inertially with some velocity
$w=H/\sqrt{H^2+1}$.
By transforming to the boundary comoving frame with constant speed,
one recovers the stationary case and a holographic interpretation
similar to the Black Hole case may be possible.
In this case the detector simply registers its ``own motion'' $w$
\cite{birrell}.
\par
However, when the Hubble constant is time-dependent or when there is
any departure from the linear Hubble law (e.g. Brane-World Cosmologies),
the response function of the detector becomes very complicated due to
long range correlations arising from non-locality
(see Section~\ref{cosmomirror}).
For the same reason, the characteristic fixed scale $M$ of a Black Hole
which introduces a lower bound on the range of quantum fluctuations
and correlation lengths probed by the detector, is lost in Cosmology.
In fact, $H(t)^{-1}$ can be as small as $\ell_p$ at early times.
In this case, the response function does not pick a $\delta$-function
projection to the surface and it thus contains memory of the parameters of
short distance physics due to the nonlocal physics at Planck scale.
These arguments lead us to speculate that the detector registers volume
degrees of freedom whenever a {\em ``nonconstant mass''} $H(t)$ is
introduced for the background.
If $H$ is time dependent from near the Planck scale then the detector
will register those features.
The motion of a ``cosmological mirror'' (the time-dependent boundary)
can be traced back in time as far as we want, up to the string scale
where short distance imprints are found.
There is no known physical reason why one should choose to stop or
start the motion of the ``cosmological mirror'' at a given
(possibly large) time rather than imposing that condition
{\em ad hoc\/}.
Therefore the uncertainty principle does not prevent the theory from
being sensitive to Planck scale physics and it allows probing of these
string scale distances.
It is not clear how or if one can apply holography for a time-dependent
boundary $H(t)$, since String Theory for time-dependent backgrounds is
not yet well understood.
The recent progress for the specific dynamic backgrounds considered in
\cite{eva} indicates that nonlocal interactions in space-time for the
boundary action are required for time-dependent backgrounds.
Accordingly, due to the correlations for the squeezed state on the
boundary, the response function (i.e.~the spectrum) will contain
information of the parameters of Nonlocal String Theory
(in space and time) giving rise to the time-dependent boundary,
i.e.~details of the boundary's past trajectory/history.
The validity of the uncertainty principle introduces a lower bound on
the range of correlations of order $\ell_p$.
When the uncertainty principle is applied to volume degrees of freedom,
it results in larger quantum fluctuations as compared to surface
fluctuations.
Therefore the chance of observing deviations from thermality in the
spectrum in Cosmology increases.
\par
The entropy change expected to occur from string interactions on the
world-sheet for time-dependent backgrounds, would indicate that there
is an energy flow between the world-sheet and the moving boundary.
This issue may become more severe in the case of non-perturbative String
Theory and would have consequences for the validity of the second law
of thermodynamics in the early universe.
Such physical arguments support the same conclusion: time-dependent
backgrounds generically would contain volume degrees of freedom, thus
they register imprints of short scale physics.
Perhaps there may be a special class of time-dependent string solutions
which may be described by local String Theory, however we are not aware
whether they would correspond to realistic cosmologies.
\par
In the absence of an Holographic Principle in Cosmology, perhaps
stochastic processes provide a relevant description of physics at
Planck scale (as discussed in Refs.~\cite{blhu}).
We do not elaborate along this direction here, although it is interesting
to explore whether a stochastic approach would change our results.
Although Planck scale physics remains beyond our grasp of understanding,
the considerations presented here are exciting and a step forward because
they indicate that the possibility of finding, in cosmological observables,
signatures of Planck scale physics and indirect evidence for String Theory,
are not far out of reach.
\begin{acknowledgments}
We would like to thank E.~Silverstein, A.~Tseytlin and W.~Unruh
for beneficial discussions.
\end{acknowledgments}
\end{document}